\documentclass[pra,twocolumn,groupedaddress,amsmath,floatfix]{revtex4}
\usepackage{graphicx}
\usepackage{bm}

\newcommand{\grad}{{\bm{\nabla}}}
\newcommand{\zh}{\hat{z}}

\newcommand{\br}{{\bf r}}
\newcommand{\be}{\begin{equation}}
\newcommand{\ee}{\end{equation}}
\newcommand{\bea}{\begin{eqnarray}}
\newcommand{\eea}{\end{eqnarray}}

\newcommand{\bv}{\bar{{\bf v}}_s}
\newcommand{\bu}{{\bf u}}

\newcommand{\rtf}{R}

\input{epsf}

\begin{document}
\title{Vortex Lattice Inhomogeneity in  Spatially Inhomogeneous Superfluids}
\author{Daniel E.~Sheehy and Leo Radzihovsky}
\affiliation{
Department of Physics, 
University of Colorado, 
Boulder, CO, 80309}
\date{February 26, 2004}
\begin{abstract}

  A trapped degenerate Bose gas exhibits superfluidity with spatially
  nonuniform superfluid density.  We show that the vortex distribution in
  such a highly inhomogeneous rotating superfluid is nevertheless
  nearly uniform. The inhomogeneity in vortex density, which diminishes
  in the rapid-rotation limit, is driven by the discrete way 
  vortices impart angular momentum to the superfluid. This effect
  favors highest vortex density in regions where the superfluid
  density is most uniform (e.g., the center of a harmonically trapped
  gas).  A striking consequence of this is that the boson velocity
   deviates from a rigid-body form exhibiting a radial-shear flow
   past the vortex lattice.

\end{abstract}
\maketitle


It has long been understood that a superfluid can only rotate by
nucleating quantized vortices~\cite{onsager,feynman}, which in turn
control most of its macroscopic properties.  In recent years, rapid
progress in the field of confined degenerate Bose gases has led to the
experimental realization of large vortex
lattices~\cite{Matthews,Madison,Haljan,Aboshaeer,Engels}.

One of the most striking features of these arrays is their apparent
{\em uniformity}, despite the strong spatial variation of the
local superfluid density imposed by the trap. These observations
cannot be simply explained by the strong vortex interaction, nor by
appealing to the imposed rigid-body rotation.  Based on purely
energetic considerations, one would expect a highly nonuniform vortex
distribution that is suppressed at the center of the trap where the
superfluid density (and therefore kinetic energy cost), as
well as vortex repulsion, are largest\cite{analogySC}. Despite some
attempts to understand this uniformity~\cite{Anglin},which has also been observed in
simulations~\cite{Feder01}, no clear physical explanation has appeared
in the literature.

In this paper we present a theory of vortices in a confined,
spatially inhomogeneous rotating superfluid.  We provide a simple
physical explanation for, and compute corrections to, a uniform vortex
array. Our main result is the vortex density $\bar{n}_v(r)$ which, in the
simplest case of a harmonic trap with a strong uniaxial anisotropy and
within the Thomas-Fermi (TF) limit, is given by
\be
\label{eq:tf}
\bar{n}_v \simeq \frac{\omega}{\pi} - \frac{1}{2\pi}
\frac{\rtf^2}{(\rtf^2 - r^2 )^2} \ln \frac{{\rm e}^{-1}}{\xi^2 \omega},
\;\;\mbox{for}\;\;\; r\ll\rtf,
\ee
with $R$ the TF radius, $\omega=\Omega m/\hbar$ the rescaled
rotational velocity $\Omega$, $m$ the boson mass, and $\xi$
the coherence length~\cite{commentTF}.  

The nearly uniform vortex distribution $\bar{n}_v\approx\omega/\pi$
 is a consequence of a balance between spatial
variations of the kinetic energy per vortex and the vortex chemical
potential, both of which scale with the local superfluid density,
$\rho_s(r)$. While it is energetically costlier to position vortices
in a region where $\rho_s(r)$ is high (the center of the trap), the
vortex chemical potential (controlled by $\rho_s(r)\omega$) is also
high there, compensating and leading to an approximately uniform vortex
density.

A spatially-dependent correction to $\bar{n}_v(r)$ in Eq.~(\ref{eq:tf})
 arises
from vortex discreteness and the related inability of the vortex state
to locally reproduce uniform vorticity corresponding to 
rigid-body rotation. In an inhomogeneous condensate, the associated
kinetic energy-density cost is spatially dependent, and is lowered
by a nonuniform vortex distribution. The reduction in $n_v(r)$ scales
%
%
with $\nabla^2 \ln \rho_s$, i.e., it is smallest 
where the 
condensate is most uniform, and leads to the strongest vortex density
suppression away from the center of the trap.  The correction vanishes
in the uniform condensate ($R\rightarrow\infty$) and, (within 
the London approximation) dense vortex (fast
rotation, $\omega\xi^2 \simeq 1$) limits, in which condensate inhomogeneity and
vortex discretness are (seemingly) unimportant~\cite{comment}. 

An immediate interesting consequence of the radial vortex lattice
distortion is that the corresponding azimuthal superfluid velocity
${\bf v}_s(r)$ deviates from a rigid-body form, exhibiting  radial-shear 
flow.  We expect that this 
will induce an azimuthal shear
distortion of the lattice, with chirality set by the sense of
the imposed rotation~\cite{commentTwist}. Below we sketch the 
derivation of these results.

A rotating superfluid is most easily analyzed in the frame in which
the boundary conditions (i.e.~the proverbial bucket) are
stationary. For experiments on trapped Bose gases, this is the
frame rotating with frequency $\Omega$ (in which the normal fluid is
stationary~\cite{feynman}).  For simplicity we focus on a trap with a
high degree of uniaxial anisotropy, which reduces the problem to
two-dimensions perpendicular to the ($z$-) axis of rotation.  
Deep in
the superfluid state the London description, which focuses on the
superfluid phase $\theta$  degree of freedom, is sufficient
and is represented by the energy
\be
E =\frac{\hbar^2}{2m} \int d^2 r\rho_s(r)[(\grad \theta)^2 
- 2\omega (\zh \times \br) \cdot\grad \theta],
\label{eq:f2}
\ee
where within the TF approximation $\rho_s(r)\approx (\mu - V(r))/g$, with
$\mu$ the boson chemical potential, $g$
the s-wave scattering potential and $V(r) = V_T(r) - \frac{1}{2} m \Omega^2r^2$
a combination of the trapping ($V_T(r)$) and \lq\lq centrifugal\rq\rq\ potentials.

Before addressing the many vortex problem, we present the solution for
a single vortex in an inhomogeneous superfluid, which, despite a
number of studies~\cite{Pismen,Lundh,Svidzinsky,Anglin02,McGee}, 
has eluded a complete solution.  For a vortex located at $\br_0$ off
the trap center, the superfluid velocity
${\bf v}_s=(\hbar/m)\grad \theta$ is determined by the Euler-Lagrange (EL)
equation for Eq.~(\ref{eq:f2}) (expressing boson number conservation)
$\grad\cdot(\rho_s(r) \grad \theta) = 0$, subject to the vorticity
constraint $\grad \times \grad \theta = 2\pi \delta^{(2)}(\br -
\br_0)$, and a boundary condition of a vanishing superflow transverse
to the boundary.  In contrast to uniform systems found in condensed
matter experiments (e.g., helium in a bucket), in an atomic trap
$\rho_s(r)$ vanishes at the boundary, thereby automatically satisfying
the boundary condition. The solution can then be expressed as
$\theta = \theta_v + \theta_a$ with $\grad \theta_v = \frac{\zh \times
(\br - \br_0)}{(\br -\br_0)^2}$ the usual (uniform-superfluid) vortex
form ensuring the topological vorticity constraint, and the
single-valued analytic phase $\theta_a$ satisfying the EL equation:
\be
\label{eq:elanalytic}
\grad\rho_s \cdot  \grad \theta_a +  \rho_s \nabla^2 \theta_a  
= - \grad \rho_s\cdot \grad \theta_v .
\ee
We reserve the full analysis of Eq.~(\ref{eq:elanalytic}) to 
Ref.~\onlinecite{Sheehy}, focusing here on main results.  Near $\br_0$, 
we find, in agreement with Refs.~\onlinecite{Pismen,Svidzinsky},
\be
\label{eq:approxgrad}
\grad \theta_a(\br) \approx  \frac{\zh \times \grad \rho_s(r_0)}{2\rho_s(r_0)}
\ln |\br - \br_0|/R.
\ee
%
%
%
Far away from the
vortex, ${\bf v}_s^a\equiv(\hbar/m)\grad \theta_a(\br) \approx 0$, vanishing
like a dipole field with negative and positive vortices at $\br_0$ and
at the center of the trap (${\bf r}=0$), respectively. Because for a
typical trap $\grad \rho_s(r_0) \propto - \hat{\bf r}_0$, and the
logarithm is negative for $\br \to \br_0$, the analytic distortion in
the superfluid velocity is perpendicular to $\br_0$ and leads to a
superflow that is no longer purely azimuthal around an off-axis
vortex. In agreement with local mass conservation, $v_s$ is smaller
on the trap-center side of the vortex (where $\rho_s(r)$ is larger)
and larger on its outside (where $\rho_s(r)$ is smaller). A refinement
of experiments~\cite{Matthews,Inouye} that have demonstrated the ability
to measure the phase variation $\theta(\phi)$ around a vortex should
allow a direct detection of the superflow distortions predicted here.

In an ideal superfluid and in the absence of other forces, a
vortex moves with the local superfluid velocity.  Therefore,
the
above result has interesting implications for vortex
dynamics. Namely, since ${\bf v}_s^a(r)$
 is finite at the center of the vortex
(where $|\br -\br_0| \approx \xi$), we find a remarkable
result: a single vortex at radius $r$, without
(sustained) externally imposed rotation will precess about the trap
center at a frequency $\omega \approx (\hbar/2m r)(\partial_r
\rho_s(r)/2\rho_s(r))\ln \frac{\xi}{R}\approx(\hbar/2m R^2)\ln
\frac{R}{\xi}$, that away from the condensate edges is roughly
independent of $r$.
Such vortex precession has in fact been seen
experimentally~\cite{Anderson}, with a quality factor of order 10.
The implication of the superflow distortion ${\bf v}_s^a(r)$ is even
richer for dynamics of a pair of vortices. We predict that two
same-charge vortices will orbit their center of charge, that will in
turn precess about the center of the trap, with two frequencies
determined by vortex separation and location relative to the axis of
the trap~\cite{Sheehy}.

We now turn to the many-vortex problem, with the goal of computing the
vortex spatial distribution in an inhomogeneous rotating superfluid.
The total ${\bf v}_s$, measured in the laboratory
frame, due to an array of $N$ vortices is the sum of the contributions
from each vortex:
\be
{\bf v}_s(\br) = \frac{\hbar}{m}\grad\theta=
\frac{\hbar}{m}\sum_{i=1}^N \frac{\zh \times (\br -\br_i)}{(\br -\br_i)^2},
\label{eq:sum}
\ee
with vortex positions $\br_i$ static in the frame of the normal
component.
In the above, we have neglected ${\bf v}_s^a(\br)$, as it has a subdominant effect
in the many-vortex problem~\cite{Sheehy}. The corresponding vortex
density is given by
$n_v(r)=(2\pi)^{-1}\grad\times\grad\theta=\sum_i^N\delta^{2}({\bf r}-\br_i)$. 

For large $\Omega$ the vortex state is dense, and we can neglect the
discrete vortex nature [embodied by Eq.~(\ref{eq:sum})] and
approximate ${\bf v}_s(\br)$ and $n_v(\br)$ by arbitrary smooth functions.  Expressing $E$ in
Eq.~(\ref{eq:f2}) in terms of ${\bf v} _s({\bf r})$ (and dropping a constant), 
we have
\be
\label{eq:f3}
E \simeq\frac{m}{2} \int d^2 r \rho_s(r)({\bf v}_s 
- \Omega (\zh \times \br))^2,
\ee
which is clearly minimized by the
rigid-body solution ${\bf v}_s = \Omega \zh \times \br$ 
corresponding to a {\it uniform\/} vortex density $\bar{n}_{v0} = \omega/\pi= m
\Omega/ \pi \hbar$.

Away from this classical rapid-rotation limit, vortex discretness begins to
matter and the above solution clearly breaks down, as ${\bf v}_s(\br)$
diverges as $1/|{\bf r}-\br_j|$ near each vortex at $\br_j$; it thus
strongly deviates from  rigid-body flow. In this regime, where a superfluid exhibits its locally
irrotational quantum nature, the summation in Eq.~(\ref{eq:sum}) can no
longer be replaced by an integration, and the minimization of 
$E$ must be done directly over the $\br_i$,
rather than over a field ${\bf v}_s(\br)$.

For a {\em uniform} infinite condensate the problem was solved long
ago by Tkachenko~\cite{Tkachenko}, who found that the solution is a
hexagonal lattice characterized by the vortex density $\bar{n}_{v0}$. Even in
the uniform case, for a finite system vortex discreteness manifests
itself in the lower-critical
rotational velocity $\Omega_{c1}\approx(\hbar/m R^2)\ln\frac{R}{\xi}$
below which no rotation is supported by the condensate.


To analyze an inhomogeneous condensate, it is essential to faithfully
incorporate vortex {\it discreteness\/} in treating the sum in
Eq.~(\ref{eq:sum}). For ${\bf r}$ near a vortex located at
$\br_j$, the flow is dominated by a diverging contribution from the
j-th vortex, with other vortices giving a subleading correction that
is smooth for $\br \to \br_j$. 
 With this observation
${\bf v}_s(\br_j+\delta{\bf r})$ near $\br_j$ is well-approximated by
\be
\label{eq:tkachenko2}
{\bf v}_s(\br_j +\delta\br)\simeq
\frac{\hbar}{m}\frac{\zh \times \delta \br}{\delta r^2} + 
\bv(\br_j),
\ee
with the smooth superflow $\bv(\br)$
\be
\label{eq:tkachenko3}
\bv(\br_j)\simeq \frac{\hbar}{m}\int d^2 r'\,
\bar{n}_v(\br') \frac{\zh\times(\br_j -\br')}{(\br_j -\br')^2},
\ee
due to all other vortices, expressed through a coarse-grained vortex
density $\bar{n}_v(\br)=\bar{n}_{v0} (1-\grad \cdot \bu(\br))$, or
equivalently vortex displacement $\bu(\br)$ that are to be
determined~\cite{Sheehy}. Interestingly, and not unlike the special
treatment of the discrete ($k=0$) ground-state BEC mode, to
retain vortex discreteness we extracted the dominant contribution to
the flow around the j-th vortex and then safely replaced the rest of the
sum by an integral over the coordinates of the other vortices.
By taking a curl with respect to $\br_j$ and solving for $\bv(\br)$,
Eq.~(\ref{eq:tkachenko3}) gives
\be
\label{eq:ve}
\bv(\br) =\Omega [ \zh
\times \br - 2\zh \times \bu(\br) ],
\ee
describing the deviation of the mean-field flow from the rigid-body
velocity due to the distortion $\bu(\br)$ from a uniform vortex lattice.

To compute the optimum vortex density
$n_v(\br)$, we express the total energy $E$ of a vortex array
[i.e., Eq.~(\ref{eq:f3})] as a sum over lattice cells\cite{baym1,Fischer}, with
each associated with a single vortex. Using
Eqs.~(\ref{eq:tkachenko2},\ref{eq:ve}) for ${\bf v}_s$ inside a cell,
making a circular-cell (of area $1/\bar{n}_v(\br_j)$) approximation,
and assuming $\rho_s(r)$ does not vary appreciably on the scale of the
vortex spacing, we compute the energy per cell. The remaining
sum over cells can be easily done, inoccuously approximating it by an
integral, i.e.,\;\;$\sum_i \to \int d^2 r \,n_v(\br) \approx \int d^2 r
\bar{n}_v(\br)$.  Thus we obtain 
\be
E\simeq \int d^2 r  \rho_s(r)\Big[ 
 \omega(1- \grad\! \cdot\! \bu) 
\ln \frac{1}{\xi^2 \omega (1-\grad\! \cdot\! \bu )}+4\omega^2 u^2\Big],
\label{eq:finf}
\ee
where we have discarded $u$-independent terms. 

Minimizing $E[{\bf u}(\br)]$ over ${\bf u}(\br)$, and re-expressing the
solution in terms of $\bar{n}_v(\br)$, we finally find
\be
\label{fineom}
\bar{n}_v(\br)  = \frac{\omega}{\pi} + \frac{1}{8\pi } \grad \Big(
\frac{1}{\rho_s(r)} \grad 
\big[ \rho_s(r)\ln \frac{{\rm e}^{-1}}{\pi \xi^2 \bar{n}_v(\br)}\big] \Big),
\ee
that gives a local vortex density in a rotated trapped superfluid,
characterized by a rotation rate $\Omega$ and local superfluid density
$\rho_s(r)$, in agreement with recent lowest Landau level 
results~\cite{comment,Ho,Watanabe,Cooper}. For a smoothly-varying $\rho_s(r)$, we expect that
vortices will have a {\it locally\/} hexagonal lattice
structure~\cite{Tkachenko}, but with a lattice parameter
$a(r)=(2/\sqrt{3}\bar{n}_v(r))^{1/2}$ that varies with radius.
Equation~(\ref{fineom}) consists of a
uniform contribution corresponding to rigid-body rotation of the
superfluid and a correction that depends crucially on the $\rho_s(r)$
profile and therefore on the shape of the trapping potential. As is
clear from its derivation, this spatial variation of the vortex
density arises from an interplay of the nonuniform trapping potential and
vortex discreteness, the latter a fundamentally quantum-mechanical
effect.

\begin{figure}[bth]
\vspace{1.4cm}
\centering
\setlength{\unitlength}{1mm}
\begin{picture}(40,40)(0,0)
\put(-50,0){\begin{picture}(0,0)(0,0)
\includegraphics{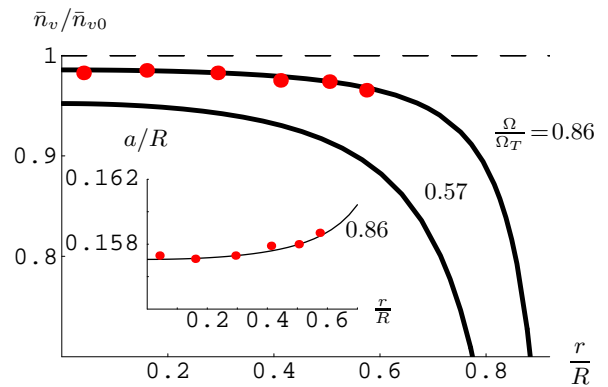}
\end{picture}}
\put(-14,50) {$\bar{n}_v/\bar{n}_{v0}$}
\put(57,4.5) {$\displaystyle \frac{r}{R}$}
\put(47,35) {$\frac{\Omega}{\Omega_T}\! =\! 0.86$}
\put(38,27) {$0.57$}
%
\put(31.0,11.5) {$\frac{r}{R}$}
\put(-2,34.0) {$a/R$}
\put(27.5,22.0) {$0.86$}
\end{picture}
\vspace{-.5cm}
\caption{(Color online) Main: Vortex density as a function of 
radius (normalized to $\bar{n}_{v0}$, dashed line) for these $\Omega$ and $R$ values (labelled by $\Omega$):
 $\Omega = .86 \Omega_T$ and $\rtf = 49 \mu m$;  $\Omega = .57 \Omega_T$
and $\rtf = 31 \mu m$; points are experimental values for the
$\Omega = .86 \Omega_T$ case adapted from Ref.~\onlinecite{JILA}.
Inset: 
Hexagonal lattice parameter $a/R$ as a function of radius for the 
$\Omega =  .86 \Omega_T$ case; points correspond to same data as in main.}
\label{fig:TFplot}\vspace{-.3cm}
\end{figure}

For a rotating condensate of size $R$, trapped in a smooth concave
potential, $\rho_s(r)$ varies on the scale $R$, leading to a {\em
negative} correction (i.e.~the second term in Eq.~(\ref{fineom})) to the
rigid-body vortex distribution, that
vanishes in the thermodynamic limit as $1/R^2$. Therefore, (as depicted in
Fig.\ref{fig:TFplot}) we predict, generically, a vortex distribution
that is denser at the center of the trap and falls off towards the
condensate edge as $\nabla\left(\nabla\rho_s(r)/\rho_s(r)\right)$.
Within our London approximation, this nonuniformity also vanishes in the fast rotation
($\omega\xi^2\simeq1$), dense-vortex limit, in which vortex cores overlap
and their discreteness is  unimportant~\cite{comment}.

Recalling that a {\em uniform} vortex distribution corresponds to a
rigid-body rotational superflow, $\bv(\br) = \Omega \zh \times
\br$, the vortex lattice distortion predicted in Eq.~(\ref{fineom})
has an immediate remarkable consequence, namely a radial shear of the
superfluid velocity, with the average rotational frequency descreasing
with radius. As a result, unlike a uniform lattice, a
radially distorted vortex lattice rotating at rate $\Omega$ cannot be
stationary with respect to the surrounding shearing
superfluid. Symmetry arguments then suggest that for a nonideal
superfluid, a vortex lattice should exhibit a chiral azimuthal radial-shear
distortion in the direction opposite to the fluid flow~\cite{commentTwist}.

We can use Eq.~(\ref{fineom}) to predict the radial vortex density
 for a rotating harmonically trapped BEC with trap frequency $\Omega_T$. Deep below
$T_c$ for a harmonic trap and away from the condensate edges we expect $\rho_s(r)$ to be
well approximated by the TF expression $\rho_s(r)\approx
\rho_s^0(1-r^2/R^2)$. Here, $R^{-2}= m (\Omega_T^2 - \Omega^2)/2\mu$ is reduced 
by the applied rotation; however, we shall regard it as an experimentally given 
parameter.
Together with a mild approximation
$\bar{n}_v(r)\approx \omega/\pi$ in the argument of the logarithm, we
immediately obtain 
Eq.~(\ref{eq:tf}), which
in Fig.~\ref{fig:TFplot} we plot for
$^{87}$Rb using 
realistic~\cite{JILA} parameters of 
 $\Omega = .86 \Omega_T$, $\rtf = 49 \mu m$ (top curve), 
and  $\Omega = .57 \Omega_T$,  $\rtf = 31 \mu m$ (bottom curve) along with data from 
Ref.~\onlinecite{JILA} for the former case.  The inset shows $a$ for the $\Omega = .86 \Omega_T$ 
case.  
Here, $\xi =\sqrt{\hbar/ m\Omega_T}$  is the TF value~\cite{Baym96} 
and  $\Omega_T = 52 s^{-1}$.  
This agreement with experimental data is achieved with 
 {\it no\/} adjustable parameters.

A combination of magnetic and optical trapping allows an unparalleled
degree of control over the single-body potential seen by the atoms. A
wide range of experimentally accessible trapping potentials allows
stringent tests of our predictions.  Consider, for example, a
sombrero potential, $V(r)$, consisting of a shallow overall trap
with frequency $\Omega_T$ and a repulsive Gaussian of width $\ell$,
with $\ell \ll R$. The  superfluid density will
display a corresponding ``dip'' near the trap center, which within the
TF approximation is given by
\be
\rho_s(r) = \rho_{s0} - \rho_{s1} \exp(-r^2/2\ell^2),
\label{eq:profile1}
\ee
with $\rho_{s0}$ ($>\rho_{s1}$) arising from the shallow part of the
trap and therefore approximately constant on the scale $\ell$.  
\begin{figure}[bth]
\vspace{1.3cm}
\centering
\setlength{\unitlength}{1mm}
\begin{picture}(40,40)(0,0)
\put(-50,0){\begin{picture}(0,0)(0,0)
\includegraphics{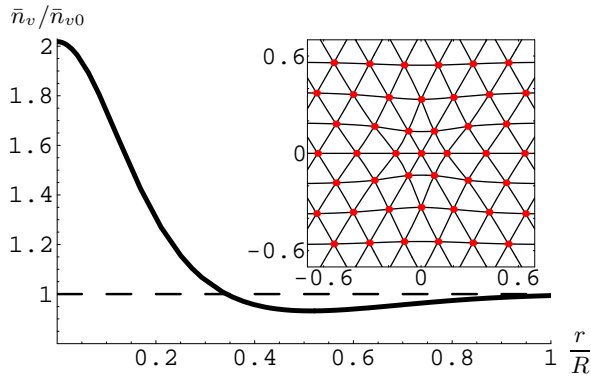}
\end{picture}}
\put(-17,49) {$\bar{n}_v/\bar{n}_{v0}$}
\put(57,4.5) {$\displaystyle \frac{r}{R}$}
\end{picture}
\vspace{-.5cm}
\caption{(Color online) Main: The solid line depicts the vortex density (normalized 
  to $\bar{n}_{v0}$; dashed line) as a function of radius for the 
  condensate profile Eq.~(\ref{eq:profile1}).   Inset: Locations of vortices
(with lines to guide the eye).
}\vspace{-.5cm}
\label{fig:distorta}
\end{figure}
Using
this $\rho_s(r)$ inside Eq.~(\ref{fineom}), we find a vortex density
profile $n_v(r)$, that is, interestingly, non-monotonic and near the
center of the trap exceeds the asymptotic rigid-body
value\cite{Sheehy}.  We plot $\bar{n}_v(r)$ for
a large, slowly rotating condensate ($R= 500 \mu m$,
$\Omega =  .2 s^{-1}$, $\rho_{s1}/\rho_{s0}=0.67$, and $\ell=.3 R$) in
Fig.~\ref{fig:distorta}.  The growth in $\bar{n}_v(r)$ near the center
of the trap manifests itself as contracted inner vortex
rings, as shown in the inset.  Unfortunately, current experiments,
which work
with relatively small condensates and vortex lattices, quickly
run out of length scales to quantitatively test details of a
spatially varying  vortex
distribution. We hope, however, that this qualitatively distinctive
vortex response to a tunable trap potential will be observable in
the next generation experiments.

To conclude, we have studied the vortex spatial distribution in a
rotating trapped superfluid and showed that, consistent with
experiments and despite a strongly inhomogeneous condensate, in the
limit of high rotation rates and a large condensate, vortex density is
nevertheless nearly uniform. We have computed the leading
spatially-dependent correction to $\bar{n}_v(r)$, which arises from an interplay
of an inhomogeneous trap potential and vortex discreteness, and showed that
{\em generically} it leads to a vortex density
that is largest at the center of the trap, in striking contrast to
simple energetic expectations. We hope that our predictions, which are
quantitatively and qualitatively consistent with recent
experiments\cite{JILA}, will stimulate more detailed experimental
work on inhomogeneous vortex states.

We gratefully acknowledge discussions with I. Coddington, E. Cornell,
P. Engels, A. Fetter and V. Schweikhard, as well as support from NSF DMR-0321848
and the Packard Foundation.
%


\begin{thebibliography}{10}
\bibitem{onsager}
L. Onsager, Nuovo Cimento, Suppl. {\bf 6}, 279 (1949).
%
\bibitem{feynman}
R.P. Feynman, in {\it Progress in Low Temperature Physics}, 
edited by C.J. Gorter (North-Holland, Amsterdam, 1955).
%
\bibitem{Matthews}
%
%
M.R. Matthews {\it et al.\/},
Phys. Rev. Lett. {\bf 83}, 2498 (1999).
\bibitem{Madison}
K.W. Madison {\it et al.\/},
Phys. Rev. Lett. {\bf 86}, 4443 (2001).
\bibitem{Aboshaeer}
J.R. Abo-Shaeer {\it et al.\/},
Science {\bf 292}, 476 (2001).
%
\bibitem{Haljan}
P.C. Haljan {\it et al.\/},
Phys. Rev. Lett. {\bf 87}, 210403 (2001).
\bibitem{Engels}
P. Engels {\it et al.\/},
Phys. Rev. Lett. {\bf 90}, 170405 (2003).
\bibitem{analogySC} The analogy with vortices in superconductors,
which are pinned by regions of suppressed superfluid density
(e.g., crystal imperfections), suggests that for Bose condensates in
a trap, vortices should be ``attracted'' to the boundary of the
trap.  The key difference is the spatial dependence of the vortex
chemical potential. 
%
%
While in superconductors it is
the {\em uniform} external magnetic field, 
%
%
in trapped superfluids
the role of vortex chemical potential is played by the product of the
rotational 
frequency $\Omega$ and the spatially varying superfluid
density $\rho_s(r)$.
\bibitem{Anglin}
J.R. Anglin and M. Crescimanno,
e-print cond-mat/0210063.
%
\bibitem{Feder01}
D.L. Feder and C.W. Clark,
Phys. Rev. Lett., {\bf 87}, 190401 (2001).
\bibitem{commentTF} The divergence of $\bar{n}_v(r)$ at $r \to \rtf$ is
due to the singular nature of the TF profile at edge of the
condensate. This arises due to the omission of gradient terms in TF
approximation of the Gross-Pitaevskii equation, which is 
invalid near $R$. Although in a more careful approximation this divergence does
not occur,
we expect that our small gradient (in $\rho_s(r)$) analysis breaks
down near the edge of the condensate where the correction to the
uniform, rigid-body rotation distribution is large.
\bibitem{comment}  
Although the vortex lattice distortion appears to vanish in the
rapid rotation ($\omega\xi^2\approx 1$) limit, this is an
artifact of our London approximation that breaks down in this dense vortex
limit. As was first shown by Ho~\cite{Ho}, in this regime a complementary
lowest Landau level (LLL) approximation applies and predicts a relation
$\bar{n}_v(r)=\omega/\pi+(4\pi)^{-1}\nabla^2\ln \rho_s(r)$, that (up to the logarithmic factor
$\propto \ln(\xi^2\omega)$) is in agreement with our prediction. In contrast to
an assumption by Ho of a uniform $\bar{n}_v$ (that leads to a Gaussian condensate
profile $\rho_s(r)$), subsequent to our work it was demonstrated~\cite{Watanabe,Cooper}
that indeed the vortex lattice distortion extends to the LLL
regime and leads to a Thomas-Fermi condensate profile in agreement
with our predictions.
\bibitem{commentTwist} We have not, however, been able to derive this
explicitly within, for example, a two-fluid hydrodynamics\cite{Sheehy}.
\bibitem{Pismen}
B.Y. Rubinstein and L.M. Pismen, 
Physica D {\bf 78}, 1 (1994).
%
%
\bibitem{Lundh}
E. Lundh and P. Ao,
Phys. Rev. A {\bf 61}, 063612 (2000).
%
\bibitem{Svidzinsky}
A.A. Svidzinsky and A.L. Fetter,
Phys. Rev. A {\bf 62}, 063617 (2000).
\bibitem{McGee}
S.A. McGee and M.J. Holland,
Phys. Rev. A {\bf 63}, 043608 (2001).
%
\bibitem{Anglin02}
J.R. Anglin, 
Phys. Rev. A {\bf 65}, 063611 (2002).
\bibitem{Sheehy}
D.E. Sheehy and L. Radzihovsky, e-print cond-mat/0406205, Phys. Rev. A (to be published).
\bibitem{Inouye}
S. Inouye {\it et al.\/},
Phys. Rev. Lett. {\bf 87}, 080402 (2001).
\bibitem{Anderson}
B.P. Anderson {\it et al.\/},
Phys. Rev. Lett. {\bf 85}, 2857 (2000).
%
\bibitem{Tkachenko}
V.K. Tkachenko, 
Zh. Eksp. Teor. Fiz {\bf 49}, 1875 (1965) 
[Sov. Phys. JETP {\bf 22}, 1282 (1966)].
\bibitem{baym1}
G. Baym and E. Chandler, J. Low Temp. Phys. {\bf 50}, 57 (1983).
\bibitem{Fischer}
U.R. Fischer and G. Baym,
Phys. Rev. Lett. {\bf 90}, 140402 (2003). 
\bibitem{Ho}
Tin-Lun Ho,
Phys. Rev. Lett. {\bf 87}, 060403 (2001).
%
\bibitem{Watanabe}
G. Watanabe, et al,
Phys. Rev. Lett. {\bf 93}, 190401 (2004).
\bibitem{Cooper}
N.R. Cooper, et al, 
Phys. Rev. A {\bf 70}, 033604 (2004).
\bibitem{JILA}
I. Coddington {\it et al.\/}, e-print cond-mat/0405240, Phys. Rev. A (to be
published).
\bibitem{Baym96} 
G. Baym and C.J. Pethick, 
Phys. Rev. Lett. {\bf 76}, 6 (1996).
\end{thebibliography}
\end{document}